# Inter-Layer Screening Length to Electric Field in Thin Graphite Film

Hisao Miyazaki[1,2], Shunsuke Odaka[1,3], Takashi Sato[4], Sho Tanaka[4], Hidenori Goto[2,4], Akinobu Kanda[2,4], Kazuhito Tsukagoshi[1,2,5*], Youiti Ootuka[4], and Yoshinobu Aoyagi[1,2,3]

[1]RIKEN, Wako, Saitama 351-0198, Japan

[2]CREST, Japan Science and Technology Agency, Kawaguchi, Saitama 332-0012, Japan

[3]Interdisciplinary Graduate School of Science and Engineering, Tokyo Institute of Technology, Midori-ku, Yokohama 226-8502, Japan

[4]Institute of Physics and Tsukuba Research Center for Interdisciplinary Materials Science, University of Tsukuba, Tsukuba, Ibaraki 305-8571, Japan

[5]Advanced Industrial Science and Technology, Higashi, Tsukuba, Ibaraki 305-8568, Japan

Electric conduction in thin graphite film was tuned by two gate electrodes to clarify how the gate electric field induces electric carriers in thin graphite. The graphite was sandwiched between two gate electrodes arranged in a top and bottom gate configuration. A scan of the top gate voltage generates a resistance peak in ambiploar response. The ambipolar peak is shifted by the bottom gate voltage, where the shift rate depends on the graphite thickness. The thickness-dependent peak shift was clarified in terms of the inter-layer screening length to the electric field in the double-gated graphite film. The screening length of 1.2 nm was experimentally obtained.

[*]E-mail address: tsuka@riken.jp

An electric conduction in thin graphite film exhibiting a high carrier mobility of up to 60,000 cm$^2$/Vs at 4 K[1] can be modulated by a gate electric field. In most experiments reported so far, the gate electric field controls density of electrons or holes. Moreover, it is also expected that the band structure of multilayer graphene can be modulated by external electric field. Especially in bilayer graphene, it is predicted that the band gap can be opened by perpendicular electric field.[2] Thus, for full control of conduction in graphene or graphite channel, it is strongly desired to understand conduction change as the gate voltage dependence.

In a thin graphite or a graphene channel, Thomas-Fermi approximation cannot be directly applied to the strongly anisotropic thin film with interlayer spacing of 0.34 nm. Furthermore, the coexistence of electrons and holes and low carrier density complicates the electric field response in the graphite film. Visscher *et al.*[3] and Guinea[4] calculated the distribution of charge in graphite under an external electric field within random phase approximation. Visscher *et al.* obtained a screening length of $\lambda = 0.54$ nm under the assumption of the existence of *intra*-layer charge polarization without *inter*-layer electron tunneling. Guinea took into account inter-layer electron tunneling in a similar model, and predicted the charge dumping in 0.7 nm from the surface with alternate charge polarity in each layer.

In this paper, we experimentally measured the interlayer screening length of thin film graphite using a configuration of a field effect transistor with top and bottom gate electrodes. We measured the resistance change for various graphite-film thicknesses as a function of the voltages applied to the two gates and analyzed the correlation between the two gate fields. We also report a simple fabrication method of top electrodes without intentional formation of a gate insulator.

We used a highly doped Si substrate with a 300-nm SiO$_2$ layer that can be used as a bottom gate electrode. Flakes of highly oriented pyrolytic graphite (HOPG) as a source material were prepared on the substrate by micromechanical cleavage method.[1] Appropriate thin film flakes were chosen by optical microscope observation.[1,5,6,7] A pair of 50-nm-thick titanium electrodes was fabricated on the graphite film as source-drain electrodes by electron beam lithography, vacuum evaporation, and the lift-off process. Subsequently, a 30-nm-thick aluminum electrode was deposited between the source-drain electrodes (Fig. 1). After exposing the device in air for several hours, the resistance between the gate electrode and the source (or drain) electrodes exceeded 100 MΩ, even though the resistance between the source-drain electrodes was maintained on the order of 1 kΩ. We suppose that an insulating layer formed between graphite film and Al electrode due to the oxidation of the Al surface in air. We use this Al electrode as a top gate electrode without an intentional insulator layer (compare refs. 8-11). In addition, after the insulation breaking due to over-voltage to the Al electrode, sufficient insulation is quickly recovered in air. In a series of the sequential measurements, the top gate voltage was operated within the non-breaking range. Thus, the recovery process did not occur during the measurement. The gate leak current was typically 10 nA which was much smaller than the current between the source and drain (~1 μA). For the low-temperature measurement, the device was immersed in liquid helium. The low-temperature measurement shows the result to be consistent with the room temperature measurement, but the noise/signal ratio is lower. The gate voltages are independently applied to the substrate bottom gate and the Al top gate. The graphite film thickness was determined based on atomic force microscope observations and the interference color in optical microscope images.[1,5,6,7]

Figure 2 shows the gate-voltage characteristic of the resistance observed in 7-nm-thick graphite film containing about 20 graphene layers. The gate voltage change of the bottom gate ($V_{bg}$) at a fixed top gate voltage ($V_{tg}$) shows an ambipolar resistance peak [Fig. 2(a)]. This ambipolar resistance peak stems from the carrier change between hole and electron. Similarly, an ambipolar resistance peak is observed as a function of $V_{tg}$ at fixed $V_{bg}$ [Figs. 2(b) and 2(c)]. The top gate voltage giving the peak, $V_{tg}^*$, depends only slightly on $V_{bg}$. In a thinner film (0.6-1 nm with 2-3 graphene layers), the resistance peak is sensitive to the bottom gate voltage $V_{bg}$ [Fig. 3(b)]. The peak $V_{tg}^*$ varies as a function of $V_{bg}$ [dotted line in Fig. 3(c)] because the electric field from each gate can reach the opposite side without being screened out.

Next, we discuss the film-thickness dependence of the ambipolar peak shift. In Fig. 4(a), the $V_{tg}^*$ shift is plotted as a ratio with respect to the $V_{bg}$ shift, $\Delta V_{tg}^* / \Delta V_{bg}$, corresponding to the slope of the dotted lines in Figs. 2(c) and 3(c). The ratio $\Delta V_{tg}^* / \Delta V_{bg}$, which was obtained for each of the seven devices, decreases exponentially as the film thickness $d$ increases. Here, we assume that the gate induced charge density decreases exponentially with the interlayer screening length $\lambda$. The model is illustrated in Fig. 4(b), where the $z$ axis is parallel to the $c$-axis of the graphite film. The charge densities induced by the bottom and top gates are expressed as $n_b(z) = n_{b0} e^{-z/\lambda}$ and $n_t(z) = n_{t0} e^{(z-d)/\lambda}$, respectively, where $n_{b0}$ and $n_{t0}$ are the charge densities on each of the surfaces. The sign of the charge density induced by the top gate is reversed in the illustration in Fig. 4(b). Since the conductivity in each graphene layer increases in proportion to the induced charge density, the two-dimensional conductivity of a layer at the $z$-position is expressed as $\sigma(z) = \sigma_0 + \alpha|n_b(z)+n_t(z)|$. Here, $\sigma_0$ is the conductivity of the neutral layer, and $\alpha$ is a constant. The total conductivity, $\sigma_{tot}$, can then be obtained by integrating over the thickness; $\sigma_{tot} = \int_0^d \sigma(z) dz/\delta = \sigma_0 d/\delta + (\alpha/\delta) \int_0^d |n_b(z)+n_t(z)| dz$, where $\delta = 0.34$ nm is the interlayer distance. The gate-dependent part in this formula is the integration in the second term, which is shown graphically as the area of the shaded portion in Fig. 4(b). For the smallest conduction at the ambipolar peak, $\frac{\partial \sigma_{tot}}{\partial n_{t0}} = \frac{\alpha}{\delta} \frac{\partial}{\partial n_{t0}} \int_0^d |n_b(z)+n_t(z)| dz = 0$ must be satisfied. Solving this equation, we obtain $n_{t0}/n_{b0} = -4/(2+e^{d/\lambda}+e^{-d/\lambda})$. Since $n_{t0}$ and $n_{b0}$ are proportional to $V_{tg}^*$ and $V_{bg}$, respectively, we can rewrite this formula as $\Delta V_{tg}^* / \Delta V_{bg} = -4a/(2+e^{d/\lambda}+e^{-d/\lambda})$, where $a$ is a constant that depends on the ratio of the capacitance of both gates. A fitting result to the data in Fig. 4(a) yields an estimated screening length of $\lambda = 1.2 \pm 0.2$ nm, under the assumption of independence of $a$ in the measured devices. The obtained screening length corresponds to 3-4 layers of graphene. This result means that the number of graphene layers should be less than 3-4 layers to make sufficient control of conduction by the gate electric field.

Ohta *et al.*[12] experimentally obtained the band structure of the graphite on SiC by angle-resolved photoemission spectroscopy. The experimental result was fitted with tight binding calculations. The interlayer charge distribution width, corresponding to the screening length in the graphite, was obtained as 0.14 and 0.19 nm for 3- and 4-layer graphene, respectively. They pointed out that their results could be an underestimation because of the intrinsic high density careers ($\sim 1 \times 10^{13}$ /cm$^2$). On the other hand, the carrier density in our samples could be less than $1 \times 10^{12}$ /cm$^2$ per single graphene sheet at the ambiploar peak.[13] The difference in the carrier density is a possible reason why there is a difference between the above estimation and our estimation.

In a theoretical prediction, the band structure of a graphite film consisting of more than 3 graphene layers has no essential difference from bulk graphite.[14] Then, in our analysis, we do not take into account the band structure change in the model, because our estimation depends mainly on nano-meter-scale films rather than the 2- or 3- layer film. This is another reason for the difference between the two screening lengths. There would be some unveiled aspects of screening length in a few layers regime in the field effect device, and further experiment in thinner films will show them, possibly related to the band structure change in the electric field.

In summary, we succeeded in making a top gate for graphite thin film using an aluminum electrode, and measured the ambipolar resistance behavior as both top and bottom gate voltage response. From the film-thickness dependence of the ambiploar behavior with respect to the scan of the two gate voltages, we determined the interlayer screening length to the external gate-electric field of the graphite to be $\lambda = 1.2 \pm 0.2 \, \text{nm}$. Thus, we can conclude that the electric conduction of graphite film thinner than 1.2 nm may be applied in high-speed operation electric switching devices.

**Acknowledgements**

This study was supported in part by a Grants-in-Aid for Scientific Research (Nos. 16GS50219, 17069004, and 18201028) from the Ministry of Education, Culture, Sports, Science and Technology of Japan. The authors would like to thank K. Novoselov and A. K. Geim for their helpful discussion on the preparation of graphene flakes.

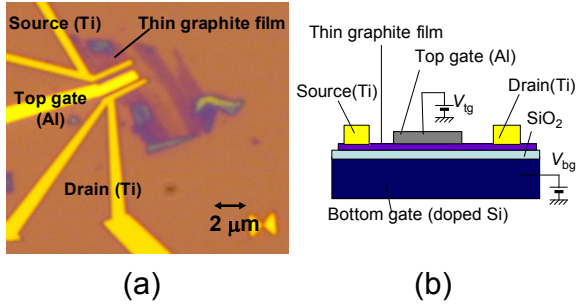
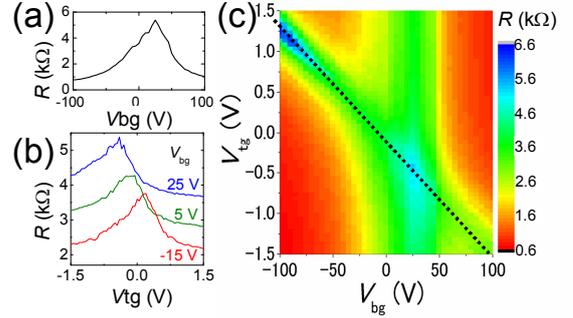

Fig. 1. (a) Optical micrograph of top gated thin graphite film with source-drain electrodes fabricated on $SiO_2$/Si substrate. (b) Schematic diagram of a dual gate graphite field effect transistor.

Fig. 3. Resistance dependence of 1-nm-thick graphite film on two gate electric fields. (a) Resistance as a function of $V_{bg}$ at $V_{tg}$ = -0.6 V. (b) Resistance as a function of $V_{tg}$ at $V_{bg}$ = -15, 5, and 25 V. (c) Color plot of resistance as a function of $V_{bg}$ and $V_{tg}$. The dotted line represents the resistance peak, and its slope is $\Delta V^*_{tg}/\Delta V_{bg}$. The measurement temperature is 4.2 K.

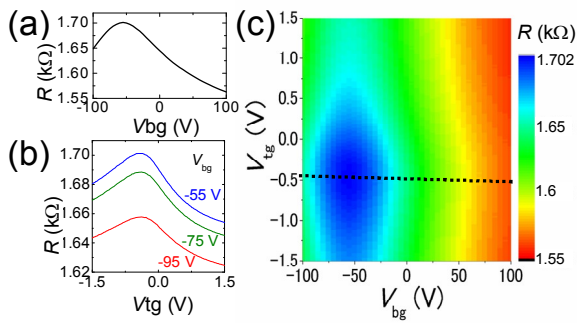
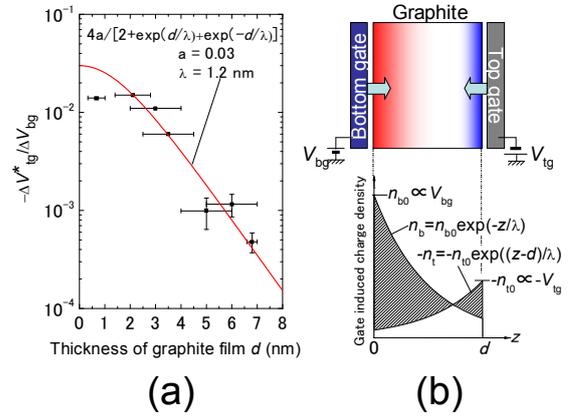

Fig. 2. Resistance dependence of 7-nm-thick graphite film on two gate electric fields. (a) Resistance as a function of $V_{bg}$ at $V_{tg}$ = -0.5 V. (b) Resistance as a function of $V_{tg}$ at $V_{bg}$ = -95, -75, and -55 V. (c) Color plot of resistance as a function of $V_{bg}$ and $V_{tg}$. The dotted line represents the resistance peak, and its slope is $\Delta V^*_{tg}/\Delta V_{bg}$. The measurement temperature is 4.2 K.

Fig. 4. (a) Relationship between the peak shift $\Delta V^*_{tg}/\Delta V_{bg}$ and the film thickness $d$ taken from seven samples. The curve is a result of the fitting based on the model in the text. (b) Schematic diagram of charge density induced by two gates. $z = 0$ and $d$ are the surface position of the bottom and top sides of the graphite film, respectively. The sign of the charge density induced by the top gate is reversed.